\documentclass[journal,twoside,web]{ieeecolor}
\usepackage{generic}
\usepackage{cite}
\usepackage{amsmath,amssymb,amsfonts}
\usepackage{algorithmic}
\usepackage{graphicx}
\usepackage{textcomp}
\usepackage{etoolbox}
\def\BibTeX{{\rm B\kern-.05em{\sc i\kern-.025em b}\kern-.08em
    T\kern-.1667em\lower.7ex\hbox{E}\kern-.125emX}}
\definecolor{g_text}{rgb}{0.0, 0.0, 0.0}
\let\mybibitem\bibitem
\renewcommand{\bibitem}[1]{%
  \ifstrequal{#1}{eriksson1996computational}
    {\color{g_text}\mybibitem{#1}}
    {  \ifstrequal{#1}{azeloglu2008heterogeneous}
        {\color{g_text}\mybibitem{#1}}
        {\ifstrequal{#1}{holzapfel2010modelling}
            {\color{g_text}\mybibitem{#1}}
            {\ifstrequal{#1}{capellini2018computational}
                {\color{g_text}\mybibitem{#1}}
                {\ifstrequal{#1}{souche2022high}
                    {\color{g_text}\mybibitem{#1}}
                {\ifstrequal{#1}{biancolini2012mesh}
                    {\color{g_text}\mybibitem{#1}}
                    {\ifstrequal{#1}{khuri2010response}
                        {\color{g_text}\mybibitem{#1}}                                  
                        {\ifstrequal{#1}{viana2010algorithm}
                            {\color{g_text}\mybibitem{#1}}                
                            {\ifstrequal{#1}{shidhore2022estimating}
                            {\color{g_text}\mybibitem{#1}}
                            {\ifstrequal{#1}{fiala2022practical}
                            {\color{g_text}\mybibitem{#1}}
                            {\ifstrequal{#1}{lin2017fluid}
                            {\color{g_text}\mybibitem{#1}}
                                {\ifstrequal{#1}{baumler2020fluid}
                                {\color{g_text}\mybibitem{#1}}
                                {\color{black}\mybibitem{#1}}}%
}                           
}                            
}
}
}
}
}
}
}
}
}
\begin{document}
\bstctlcite{IEEEexample:BSTcontrol}
\title{Calibration of the mechanical boundary conditions for a patient-specific thoracic aorta model including the heart motion effect}

\author{Leonardo Geronzi, Aline Bel-Brunon, Antonio Martinez, Michel Rochette, Marco Sensale, Olivier Bouchot, Alain Lalande, Siyu Lin, Pier Paolo Valentini,  Marco Evangelos Biancolini
\thanks{Manuscript received on July xx, 2022;  this work has received funding from the European Union’s Horizon 2020 programme under the MarieSkłodowska-Curie grant agreement No 859836, MeDiTATe.}
\thanks{Corresponding author: L. Geronzi (leonardo.geronzi@uniroma2.it)}
\thanks{L. Geronzi, A. Martinez, P.P. Valentini and M.E. Biancolini are with Department of Enterprise Engineering ”Mario Lucertini”, University of Rome Tor Vergata, Roma, Italy }
\thanks{A. Bel-Brunon is with INSA Lyon, LaMCoS, Lyon, France}
\thanks{L. Geronzi, A. Martinez, M. Rochette, M. Sensale  are with Ansys France, 69100, Villeurbanne, France}
\thanks{A. Lalande is with Medical Imaging Department, University Hospital of Dijon, 2100 Dijon, France, and with ImViA Research Laboratory}
\thanks{Siyu Lin is with Imaging and Artificial Vision Research Laboratory (ImViA), Dijon, France and University of Burgundy, Dijon, France}
\thanks{O. Bouchot, is with department of cardiovascular and thoracic surgery, University Hospital of Dijon, Dijon, France and with ImViA Research Laboratory, Dijon, France}}
\maketitle

\begin{abstract}

Objective: we propose a procedure for calibrating 4 parameters governing the mechanical boundary conditions (BCs)  of a thoracic aorta (TA) model derived from one patient with ascending aortic aneurysm. The BCs reproduce the visco-elastic structural support provided by the soft tissue and the spine and allow for the inclusion of the heart motion effect.

Methods: we first segment the TA from magnetic resonance imaging (MRI) angiography and derive the heart motion by tracking the aortic annulus from cine-MRI. A rigid-wall fluid-dynamic simulation is performed to derive the time-varying wall pressure field. We build the finite element model considering patient-specific material properties and imposing the derived pressure field and the motion at the annulus boundary. The calibration, which involves the zero-pressure state computation, is based on purely structural simulations.  After obtaining the vessel boundaries from the cine-MRI sequences, an iterative procedure is performed to minimize the distance between them and the corresponding boundaries derived from the deformed structural model. A strongly-coupled fluid-structure interaction (FSI) analysis is finally performed with the tuned parameters and compared to the purely structural simulation.

Results and Conclusion: the calibration with structural simulations allows to reduce maximum and mean distances between image-derived and simulation-derived boundaries from 8.64 mm to 6.37 mm and from 2.24 mm to 1.83 mm, respectively. The maximum root mean square error between the deformed structural and FSI surface meshes is  0.19 mm. This procedure may prove crucial for increasing the model fidelity in replicating the real aortic root kinematics.

\end{abstract}

\begin{IEEEkeywords}
Ascending aortic aneurysm, boundary conditions, heart motion, calibration, soft tissue
\end{IEEEkeywords}

\section{Introduction}
\label{sec:introduction}

\IEEEPARstart{C}{ardiovascular} patient-specific modeling aims at bringing a new insight into biomedical issues with the purpose of improving diagnosis, optimizing clinical treatments and predicting possible surgical outcomes \cite{neal2010current}.
Generally speaking, a  patient-specific finite element (FE) model should incorporate the discretized geometry, the governing equations, the initial state information and the boundary conditions (BCs) \cite{vidal2012tuning}. 
Its accuracy in reproducing the behaviour of the real vessel should be evaluated to confirm the ability to predict possible outcomes of clinical interest \cite{antiga2008image}. This is notably true for pathologies such as ascending aortic aneurysm (AsAA) and aortic dissection, for which there is currently interest in identifying accurate biomarkers to understand and predict the possible evolution of the disease \cite{martin2015patient,elefteriades2010thoracic,guo2018}. 

\begin{figure*}[ht]
\centering
\includegraphics[width=175mm]{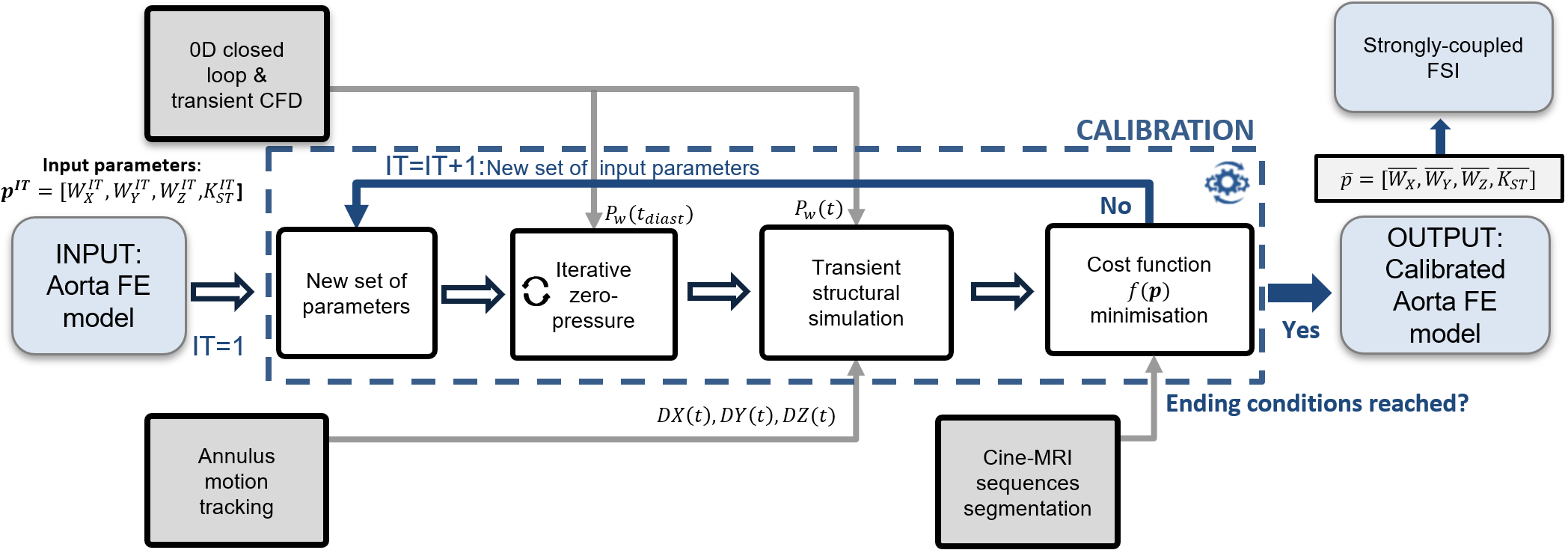}
\caption{Workflow of the calibration procedure.}
\label{fig1}
\end{figure*}

AsAA rupture usually occurs when the mechanical stresses acting on the aneurysm exceed the strength of the degenerated aortic wall \cite{vorp2003effect}. In the last few years, attention has turned towards bio-mechanical analyses to better understand the underlying mechanisms governing AsAA pathogenesis. Parameters like the wall stress cannot in fact be directly measured but can be determined computationally \cite{celi2014three}. Computational biomechanics and finite element analysis (FEA) represent valid techniques to investigate mechanical parameters and evaluate the risk associated with the pathology \cite{doyle2016computational,pasta2012effect,wolters2005patient}.

A deep understanding of the phenomena related to the evolution of the AsAA disease requires high-fidelity simulations of the human vasculature based on patient-specific geometries and accurate biomechanical models. To build them, a workflow in which reconstructions from medical images are integrated with experimental data is usually required \cite{redaelli2020cardiovascular}. The high fidelity introduced by considering the aortic root motion and its role in increasing the wall stress and strain has been shown in \cite{beller2004role} where Beller et al. proved how the downward  displacement of the aortic root increased the stress on the wall and is potentially able to adversely affect the development of aortic dissection.  Wittek et al.  \cite{wittek2016cyclic} compared the three-dimensional time-varing wall kinematics of the ascending and the abdominal aorta: the AsAA undergoes a complex deformation with alternating clockwise and counterclockwise twist. This study hence revealed the key role of heart motion that stretches the aortic root and increases AsAA wall tension \cite{cutugno2021patient}.  The importance of taking cardiac motion into account when assessing fluid dynamic outputs for the evaluation of patient status was presented in \cite{Wendell}. Weber et al. \cite{weber2009heartbeat} analysed the aortic displacement due to heartbeat in patients with chronic type B aortic dissection showing how it was substantially higher for the ascending aorta compared to the aortic arch and the descending tract.
Rueckert et al. \cite{rueckert1997automatic} introduced a technique that employed a deformable model and an energy minimization approach to track specific thoracic aorta parts.

When analyzing the wall deformation, it appears essential to integrate the patient-specific tissue material properties \cite{silver1989mechanical}, include the pre-stress of the vessel \cite{liu1988zero} and consider its interaction with the structures around it \cite {taylor2009patient,reymond2013physiological}. In \cite{petterson2019including}, the impact of considering the tissues surrounding the abdominal aorta and the presence of the spine on the stress at the wall has been investigated by integrating information derived from ultrasound images. An interesting example of external tissue support is presented in \cite{strocchi2020simulating} for the heart muscle. Gindre et al. \cite{gindre2016patient} proposed a model of abdominal aorta with viscoelastic external tissue support defined over the entire wall. 
Moireau et al. \cite{moireau2012external} and then Baumler et al. \cite{baumler2020fluid}, without considering the presence of the valve and removing the Valsalva sinuses, introduced a boundary condition (BC) along the thoracic aorta (TA) wall that consisted of a viscoelastic term representing the support provided by the surrounding tissues and organs. The external tissue support on the outer arterial wall was introduced through Robin (Fourier-type) boundary conditions \cite{eriksson1996computational}, dividing the aorta into a few large macro-areas. In \cite{moireau2013sequential}, exploiting multi-phase CT, a first approach with a sequential method to calibrate the mechanical boundary condition for the model proposed in \cite{moireau2012external}  was presented. 

In this paper, we  propose a method to calibrate 4 parameters governing the wall boundary conditions of a patient-specific thoracic aorta model accounting for the annulus motion. Our approach incorporates visco-elastic mechanical BCs to represent the support provided by surrounding soft tissue and the interaction of the aorta with the spine.  The fluid-dynamic model, built to derive the pressure field at the wall, which also involves the whole area of the Valsalva sinuses and includes the modelling of the aortic valve, is coupled with a 0D closed loop reproducing the full cardiovascular circulation. Through an iterative  procedure based on purely structural simulations, we aim at increasing the model fidelity tuning the parameters governing the mechanical BCs with the goal of obtaining an improved correspondence between simulated and image-derived displacements. In each iteration, starting from the derived unloaded state and applying the pressure load previously derived, 4 cardiac cycles are reproduced. A strongly-coupled fluid-structure interaction (FSI) analysis is finally performed with the BCs controlled by the calibrated parameters and the results are compared to the ones derived from purely structural simulation.

\section{Materials and Methods}
The evaluation of bio-mechanical parameters, such as wall stress and strain should ideally be performed considering the blood flow pressure effects on the aortic wall \cite{brown2012accuracy}. In order to correctly model this phenomenon, we decided to take these deformations due to flow ejection into account in the calibration by proposing a decoupled fluid-structure interaction method in which we performed a single computational fluid dynamic (CFD) to derive the pressure field at the wall and imposing this load for several structural simulations \cite{chandra2013fluid}.

An overview of the entire procedure, applied here to a patient with AsAA, is shown in Fig. \ref{fig1}. It consists of two steps: (1) a preparatory step in which we built the thoracic aorta FE model with mechanical boundary conditions controlled by non-calibrated initial parameters taken from Gindre et al. \cite{gindre2016patient}. A preliminary transient CFD simulation coupled with a 0D closed loop representing the cardio-vascular system (CVS) and used to impose quasi patient-specific 0D-derived boundary conditions was performed to obtain the wall pressure field $P_w(t)$. In parallel, the motion of the annulus was tracked from cine-MRI and its global displacement in the three spatial directions  ($DX(t)$, $DY(t)$ and $DZ(t)$) was derived. The splines representing the boundaries of the aorta were extracted and the alignment between them and the FE model was ensured through  Iterative Closest Point.
(2) Inside the calibration loop, first the steady-state zero-pressure model was derived by solving an inverse problem. Then, starting from this new unloaded configuration, a structural simulation was performed reproducing 4 cardiac cycles, applying $P_w(t)$ and the displacement retrieved from the annulus motion. The intersections between the deformed FE model and the cine-MRI planes were computed and compared to the splines previously obtained from the segmentation of the 2D dataset. This iterative procedure was repeated until the convergence criteria described later in this paper were reached. In this way, the value of the parameters at the end of the calibration was the most suitable to reproduce the aorta displacement derivable from cine-MRI data. The sensitivity of the method was tested by performing the calibration again with 3 different initial guesses chosen after inspecting the parameters space through a Response Surface \cite{khuri2010response}. Once the model was tuned, we performed a fully-coupled FSI to verify the differences with the purely structural simulation. The steps of the procedure are detailed in the following paragraphs.

\subsection{Image dataset and experimental data}

This work was based on retrospective and experimental data collected from a patient who underwent surgery for AsAA. The patient, a 67-year-old man, 184 cm tall and weighing 89 kg, presented a maximum ascending aorta diameter of 52 mm and a stenotic bicuspid aortic valve type 1 L-R, the most common congenital cardiac abnormality associated with AsAA \cite{losenno2012bicuspid}. Non-invasive measurement during the MRI acquisition revealed diastolic and systolic pressure respectively of 72 mmHg and 105 mmHg at the level of the arm.
The proposed approach relies on cine-MRI sequences to derive the aorta kinematics benefiting from their good temporal resolution and excellent signal-to-noise ratio \cite{link1993magnetic}. The MRI acquisition was conducted 6 days before surgery using a 3 Tesla scanner (Siemens Healthineers, Erlangen, Germany)  with a phased thoracic coil. The acquisition was performed in apnea and the patient’s heart rate during the procedure was 61 bpm. The cine-MRI sequence, consisting of 256x208 pixel images, was electrocardiogram-gated (ECG-gated), providing 25 temporal frames of the cardiac cycle with a 2D spatial resolution of 1.5 mm x 1.5 mm, a slice thickness of 5 mm and a temporal resolution of 42 ms according to the patient's cardiac rhythm. In the same exam, magnetic resonance angiography (MRA) in breath-holding conditions was performed. The ECG-gated MRA images were acquired at the end-diastolic phase and their spatial resolution was 1 mm x 1 mm x 1 mm. This dataset consisted of 416x312x72 voxel images. 

\begin{figure}[!h]
\centering
\includegraphics[width=80mm]{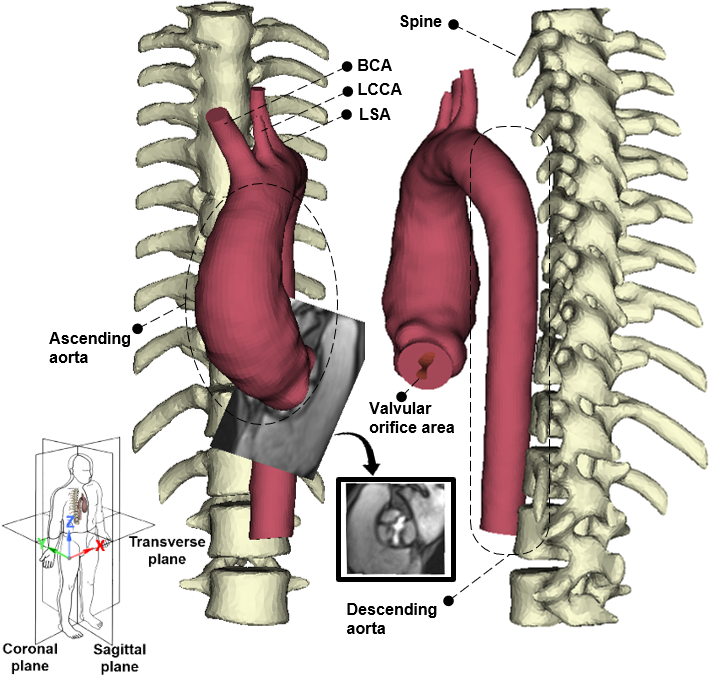}
\caption{The aortic model, the spine and the cine-MRI slice from which the area corresponding to the valve jet could be observed are shown. The fluid domain inlet was created on the specific plane by segmenting the area extracted from this cine-MRI acquisition during systole.}
\label{fig2}
\end{figure}
\subsection{Segmentation and mesh generation}

9 sagittal and 2 oblique Left Ventricular Outflow Tract (LVOT) cine-MRI acquisitions from which the aorta displacement and deformation could be detected were segmented in 3D Slicer \cite{kikinis20143d} by region growing and subsequent manual refinement.  From each mask obtained, the boundaries corresponding to the intersection of the cine-MRI plane and the TA were obtained using VTK Canny Edge Detector filters.  The splines built from the set of points derived from the centroids of the pixels belonging to the aortic boundaries were reconstructed and relaxed with 20 iterations of Taubin Smoothing.   It is important to note that conventional cine-MRI can only capture the motion in a 2D plane and the vessel portion obtained at each frame corresponded to a different part of the aorta. For each time frame, all the points generating the splines on the 11 planes resulted in a point cloud. The first frame of the cine-MRI sequence was identified at the end of diastole, corresponding to the cardiac phase for which the MRA was performed. To avoid an ill-conditioned problem, we calibrated the mechanical BCs studying the motion only in a subset of the 25 frames. In particular, assuming the point cloud derived from the first frame as initial reference, the subsequent frame with a nearest neighbour distance \cite{clark1954distance} between the respective point cloud and the reference one higher than 1.5 mm was taken as new reference and involved in the calibration, ensuring the inclusion of boundaries for which a displacement of at least 2 pixels occurred. Thus, the calibration involved the frames $\varphi$=\{3,5,7,9,11,13,19\}.

The anatomical models shown in Fig. \ref{fig2} and representing the aorta and spine were obtained from the MRA images.
The three-dimensional MRA dataset was segmented using a thresholding method. The aortic geometry was imported in ANSA pre-processor (BETA CAE Systems, Switzerland) to create the computational grid. The structural mesh for the aortic wall $\Gamma_{W}$ was built using 14000 fully-integrated quadrilateral shell elements and the ascending portion of the aorta was identified and isolated from the rest to perform the calibration.
The CFD grid consisted of 2 million of structured hexahedral elements. Eight inflation layers \cite{gross2019mesh} for a total thickness of 2 mm with a growth rate equal to 1.5 were generated. Using a specific cine-MRI acquisition perpendicular to the vessel centerline and at the level of the valve leaflets, the inlet of the aorta was obtained by segmenting the 2D area where the valvular orifice was visible (Fig. \ref{fig2}) \cite{oliveira2019bicuspid,bonomi2015influence}.

To account for potential variations in diaphragm position between subsequent apnea periods, a rigid registration was conducted to achieve spatial alignment between the 3D aortic model and the point cloud derived from the first frame of the cine-MRI sequence. The registration process involved 10 steps of Iterative Closest Point to ensure accurate alignment.

\subsection{Annulus tracking}
\begin{figure}
\centering
\includegraphics[width=85mm]{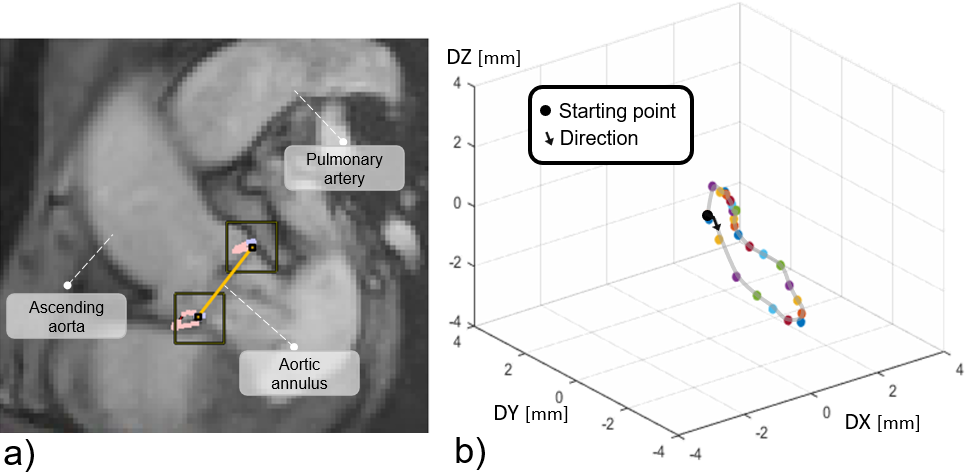}
\caption{a) The annulus plane tracked in a cine-MRI sequence. b) The motion ($DX(t), DY(t), DZ(t)$) derived from  all landmarks.}
\label{fig4}
\end{figure}

The aortic annulus was tracked through anatomical landmarks using a semi-automatic Python algorithm in Blender \cite{blain2019complete}. To derive the motion in the three directions of space, the 2 oblique LVOT slices and 5 of the 9 sagittal slices containing the annulus were used for the tracking. For the first frame of every sequence, a landmark consisting of a window pattern area of 10 x 10 pixels was inserted at each annulus extremity.  The search area was defined through a second larger window of 15 x 15 pixels. Starting with the first frame, in which the initial landmarks were manually placed, a planar tracking was performed for the subsequent frames by correlating grey levels within the new search area with the pattern area of the frame before. At each step, the reference frame was updated with the newly identified window (Fig. \ref{fig4}a). A successful tracking was accepted when the maximum correlation between the grey level of the newly positioned window pattern and that of its position at the frame before was achieved. Subsequently, the 2D coordinates of the landmarks in the images were traced back to the 3D positions using the DICOM information. To capture the cardiac muscle motion and mitigate the influence of blood flow pressure-induced wall dilation  \cite{sucha2015does}, the average of all the extracted landmark displacements was obtained for each frame. In this calculation, the contribution in the x,y, and z directions of each landmark belonging to a cine-MRI plane, of which $\mathbf{n}=[n_x, n_y, n_z]$ was its normal, was respectively weighted by $w_x=1/(1-n_x^2)^{\frac{1}{2}}$, $w_y=1/(1-n_y^2)^{\frac{1}{2}}$ and $w_z=1/(1-n_z^2)^{\frac{1}{2}}$ to scale the value considering the orientation of the plane to which the landmark belonged. In case of cine-MRI plane perpendicular to one of the 3 main axes, the undefined weight relative to it was assumed as zero and the null displacement value was excluded for the average computation. A unique 3D  motion ($DX(t), DY(t), DZ(t)$) was thus obtained (Fig. \ref{fig4}b). 

\subsection{Fluid dynamic model and 0D closed loop}

\begin{table}
\centering
\caption{0D lumped parameters}
\label{table}
\setlength{\tabcolsep}{3pt}
\begin{tabular}{|p{35pt}|p{100pt}|p{95pt}|}
\hline
Symbol& 
Name& 
Value \\
\hline
\vspace{0.0005 in}$C_{BCA}$ & \vspace{0.0005 in}Capacitance after BCA & \vspace{0.0005 in}$1.1\times10^{-9}$  m$^{2}$/Pa \\
$R_{BCA}$ & Resistance after BCA & $5.1\times10^{7}$ (Pa$\cdot$s)/m$^{2}$\\ 
$L_{BCA}$ & Inductance after BCA  & $1.2\times10^{3}$ (Pa$\cdot$s$^{2}$)/m$^{3}$\\
\vspace{0.0005 in}$C_{LCCA}$ & \vspace{0.0005 in}Capacitance after LCCA & \vspace{0.0005 in}$1.2\times10^{-10} $m$^{2}$/Pa \\
$R_{LCCA}$ & Resistance after LCCA & $2.6\times10^{7}$ (Pa$\cdot$s)/m$^{2}$\\ 
$L_{LCCA}$ & Inductance after LCCA  & $1.9\times10^{7}$ (Pa$\cdot$s$^{2}$)/m$^{3}$\\
\vspace{0.0005 in}$C_{LSA}$ & \vspace{0.0005 in}Capacitance after LSA & \vspace{0.0005 in}$1.0\times10^{-10}$  m$^{2}$/Pa \\
$R_{LSA}$ & Resistance after LSA & $1.7\times10^{7}$ (Pa$\cdot$s)/m$^{2}$\\  
$L_{LSA}$ & Inductance after LSA  & $1.0\times10^{4} $(Pa$\cdot$s$^{2}$)/m$^{3}$\\
\vspace{0.0005 in}$C_{DA}$ & \vspace{0.0005 in}Capacitance after DA & \vspace{0.0005 in}$9.8\times10^{-9} $m$^{2}$/Pa \\
$R_{DA}$ & Resistance after DA & $1.9\times10^{7} $(Pa$\cdot$s)/m$^{2}$\\ 
$L_{DA}$ & Inductance after DA  & $1.4\times10^{6} $ (Pa$\cdot$s$^{2}$)/m$^{3}$\\
\vspace{0.0005 in}$R_{D1}$ & \vspace{0.0005 in}Resistance 1 Downstream & \vspace{0.0005 in}$1.8\times10^{9} $ (Pa$\cdot$s)/m$^{2}$\\ 
$R_{D2}$ & Resistance 2 Downstream & $1.1\times10^{8}$ (Pa$\cdot$s)/m$^{2}$\\
$C_{D}$ & Capacitance Downstream  & $3.7\times10^{-9}$  m$^{2}$/Pa \\
\vspace{0.0005 in}$R_{U1}$ & \vspace{0.0005 in}Resistance 1 Upstream & \vspace{0.0005 in}$2.4\times10^{8}$ (Pa$\cdot$s)/m$^{2}$\\ 
$R_{U2}$ & Resistance 2 Upstream & $7.1\times10^{7} $ (Pa$\cdot$s)/m$^{2}$\\ 
$C_{U}$ & Capacitance Upstream  & $1.7\times10^{-10}$ m$^{2}$/Pa \\
\hline
\end{tabular}
\label{tab1}
\end{table}

\begin{figure*}[!h]
\centering
\includegraphics[width=165mm]{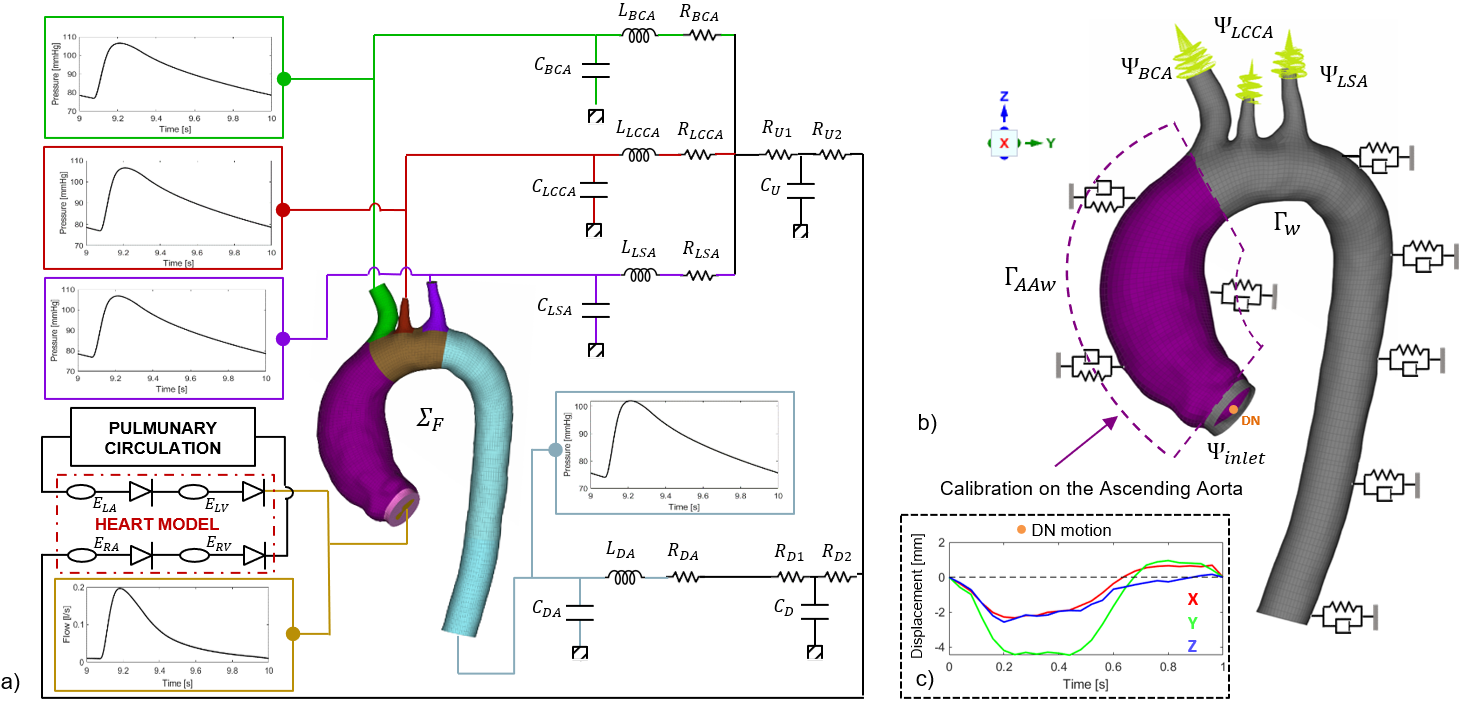}
\caption{a) The CFD model of the aorta and the lumped parameters used to generate the boundary conditions. b) The structural model with the Robin BCs: the ascending tract on which the calibration is performed is shown in purple. c) The dummy node displacement.}
\label{fig3}
\end{figure*}

The CFD boundary conditions were derived using a quasi patient-specific 0D closed loop reproducing the CVS \cite{shi2011review}. The lumped parameters of the systemic circulation, shown in Fig. \ref{fig3}a and whose values are reported in Table \ref{tab1}, were tuned in order to match the cuff pressure measurement using the method proposed in \cite{tomasi2020patient,romarowski2018patient}.
The time-varying elastances of the heart chambers were modeled using the characteristic elastances for the right and left ventricle and atrium ($E_{RV}$, $E_{LV}$, $E_{RA}$, $E_{LA}$) and the activation functions proposed in \cite{blanco20103d} and \cite{bozkurt2019mathematical} and adapted by considering peak systolic time $T_{PS}=0.28$ s, ventricular relaxation time $T_{VR}=0.45$ s and duration of the cardiac cycle $T=0.98$ s. The pulmonary circulation parameters were taken from literature \cite{korakianitis2006concentrated}. 

For the 3D fluid-dynamic domain $\Sigma_{F}$, the flow model selected was Viscous–Laminar with an incompressible and Newtonian fluid. Blood density $\rho=1100$ kg/m$^3$ and  dynamic viscosity $\mu = 4$ cP were obtained from literature \cite{cokelet2007macro}. A no-slip boundary condition was set to the wall. The rigid-wall CFD simulation was performed in Ansys\textsuperscript{\textregistered} Fluent v22.1 (Ansys Inc., Pittsburgh, PA, USA) solving the following Navier-Stokes equations:

\begin{equation}
\label{NS_eq}
\centering
\frac {\textup{D}\mathbf{u}}{\textup{D}t}= - \frac{1}{\rho }\triangledown P +\nu \triangledown^2 \mathbf{u}
\end{equation}

\begin{equation}
\label{NS_eq2}
\centering
  \triangledown \cdot \mathbf{u}=0
\end{equation}
with $\mathbf{u}$ the velocity, $P$ the pressure and $\nu$ the kinematic viscosity of the fluid.
10 cardiac cycles were simulated to achieve cycle-independent results. The time-varying wall pressure field $P_w(t)$ of the last cardiac cycle was stored.

\subsection{Structural Model}
The structural model is shown in Fig. \ref{fig3}b.
Experimental tests were performed immediately after surgery to derive the patient's ex-vivo material properties for the arterial wall\cite{lin2019local}. Bi-axial tests were conducted on eight samples from different regions of the ascending aorta, resulting in eight stress-strain curves and thickness measurements. For the FE model, the average thickness of 1.3 mm was used and an average stress-strain curve was computed from the experimental data. We introduced a non-linear isotropic strain-energy potential W for a 3-parameter hyperelastic Mooney-Rivlin model \cite{wex2015isotropic}, expressed as function of the Cauchy invariant $\bar{I_1}$ and $\bar{I_2}$ \cite{kim2012comparison}:
\begin{equation}
W= C_{10}\left(\bar{I_1}-3\right)+C_{01}\left(\bar{I_2}-3\right)+C_{11}\left(\bar{I_1}-3\right)\left(\bar{I_2}-3\right)
\label{MR_material}
\end{equation}
where $C_{10}$=-1.47 MPa, $C_{01}$=1.62 MPa and $C_{11}$=2.31 MPa were derived from a least-square fitting of the averaged curve.

To represent the relationship between aorta and spine and neglecting the intra-thoracic pressure as the acquisition was performed in breath-hold \cite{gindre2016patient}, the following equation modelling the external tissue support  through Robin BCs was used:

\begin{equation}
\boldsymbol{\mathrm{\sigma_{ext}}} = \boldsymbol{\mathrm{-K}} \boldsymbol{\mathrm{x}} -{\boldsymbol{\eta}} \boldsymbol{\mathrm{\dot{x}}}
\label{tissue_support}
\end{equation}
with $\boldsymbol{\mathrm{\sigma_{ext}}}$ the stress due to the external load, $\mathbf{K}$ and ${\boldsymbol{\eta}}$ the parameters that model the  elastic and viscous tissue support, $\boldsymbol{\mathrm{x}}$ the displacement, $\boldsymbol{\mathrm{\dot{x}}}$ the velocity. $\mathbf{K}$ and ${\boldsymbol{\eta}}$ can be decomposed along the three anatomical directions: $x$-coronal, $y$-transvers, $z$-sagittal. 
More in detail, we connected each node $i \in {\Gamma_{W}}$ to three dampers and three springs, each for a $j$-direction of the 3D space. 
The stiffness of each spring was given by:
\begin{equation}
K_{j_{i}} = K_{ST} + W_{d_{i}}W_j K_{SPINE}
\label{stiffness}
\end{equation}
where $K_{ST}$ was the first sought parameter representing the stiffness of the soft tissue around the aorta. We assumed in this case the force applied by the soft tissue equal in the three spatial directions. $W_{j}$ was the weight to be optimised for each direction of the space; $K_{SPINE}=10^6$ Pa/m, assessed from \cite{gindre2016patient}, was the stiffness that related the aorta to the spine and $W_{d_{i}}$ the scaling factor related to the distance to the vertebrae:
\begin{equation}
W_{d_{i}}= 1-\alpha_s \frac{d_i}{d_{max}}
\label{weights}
\end{equation}
with $\alpha_s$= 0.95, $\boldsymbol{d}$ the vector containing the minimum Euclidean distance of each TA node from the spine and $d_{max}$ the maximum distance, in this model equal to 142 mm. For the brachiocephalic artery (BCA), left common carotid artery (LCCA) and left subclavian artery (LSA), an additional set of springs (in green in Fig. \ref{fig3}b) was added to reproduce the influence of the upper trunk vasculature on the aortic vessel wall \cite{liu2007surrounding}. The nodes of the boundaries $\Psi_{BCA}$, $\Psi_{LCCA}$ and $\Psi_{LSA}$ were respectively connected to three new additional nodes located along their outgoing central axis at a distance of twice the mean diameter. The upstream vasculature stiffness $K_{US}$ on each boundary was assumed the same for every 2D element and defined by the following arithmetic mean:
\begin{equation}
\left\{\begin{matrix}
K_{US_{BCA}}=\mathrm{mean}(\boldsymbol{\mathrm{K_{\Psi_{BCA}}}}) \\ 
K_{US_{LCCA}}=\mathrm{mean}(\boldsymbol{\mathrm{K_{\Psi_{LCCA}}}}) \\ 
K_{US_{LSA}}=\mathrm{mean}(\boldsymbol{\mathrm{K_{\Psi_{LSA}}}})
\end{matrix}\right.
\label{Upstream}
\end{equation}
where $\boldsymbol{\mathrm{K_{\Psi_{BCA}}}}$, $\boldsymbol{\mathrm{K_{\Psi_{LCCA}}}}$, and $\boldsymbol{\mathrm{K_{\Psi_{LSA}}}}$ denote the vector containing the stiffnesses of the subgroup of springs belonging to $\Psi_{BCA}$, $\Psi_{LCCA}$, and $\Psi_{LSA}$, respectively. $\boldsymbol{\eta}$ is assumed constant and equal to 10$^5$ (Pa$\cdot$s)/m for each node and direction in space \cite{moireau2012external}.
The time-varying pressure field $P_w(t)$ obtained from the CFD was applied to the aortic wall $\Gamma_{W}$. Since a Dirichlet boundary condition, such as the displacement resulting from the sequential images, was not physically compatible with the pressure load, a strategy involving a dummy node (DN) was introduced. We used the following constraint equations to control the motion of the inlet boundary $\Psi_{inlet}$:

\begin{equation}
\left\{\begin{matrix}
n_{\Psi_{inlet}} dx_{DN}(t) - \sum_{1}^{n_{\Psi_{inlet}}}dx_i(t) = 0  \\
n_{\Psi_{inlet}}  dy_{DN}(t) - \sum_{1}^{n_{\Psi_{inlet}}}dy_i(t) = 0  \\
n_{\Psi_{inlet}} dz_{DN}(t) - \sum_{1}^{n_{\Psi_{inlet}}}dz_i(t) = 0 
\end{matrix}\right.
\label{dummy}
\end{equation}
where $n_{\Psi_{inlet}}$ was the number of nodes of $\Psi_{inlet}$, $dx_{DN}(t), dy_{DN}(t), dz_{DN}(t)$ the $x$, $y$ and $z$ displacements imposed to DN, shown in Fig. \ref{fig3}c and respectively set equal to $DX(t)$, $DY(t)$ and $DZ(t)$, and $dx_{i}(t), dy_{i}(t)$ and $ dz_{i}(t)$ the displacement of a single node of $\Psi_{inlet}$. This choice ensured that the inlet could deform during the cardiac cycle. 
Notably, we neglected the effects of rotation of the aorta around the axis passing through the centre of the annulus as it was not possible to retrieve this information from 2D images.

\subsection{Zero-pressure computation} 
The geometry  obtained from MRA corresponded to the loaded diastolic state. To include the pre-stress, the zero-pressure configuration \cite{bols2013computational} was derived solving an inverse problem based on the iterative method reported in \cite{rausch2017augmented}. From the end-diastolic phase of the rigid-wall CFD simulation, the static pressure $P_w(t_{diast})$ on ${\Gamma_{W}}$ was derived and mapped on the structural model. The aorta was only constrained by the entire set of springs as the dampers have no effect in steady-state analysis. For the convergence criterion of the iterative algorithm, an error on the maximum euclidean nodal distance (threshold of 0.1 mm) between the pressurized zero-pressure model and the one derived from MRA in diastole was included and a limit of 20 iterations was established. The Ansys LS-Dyna implicit solver was used.

\subsection{The calibration method}
\begin{figure}
\centering
\includegraphics[width=60mm]{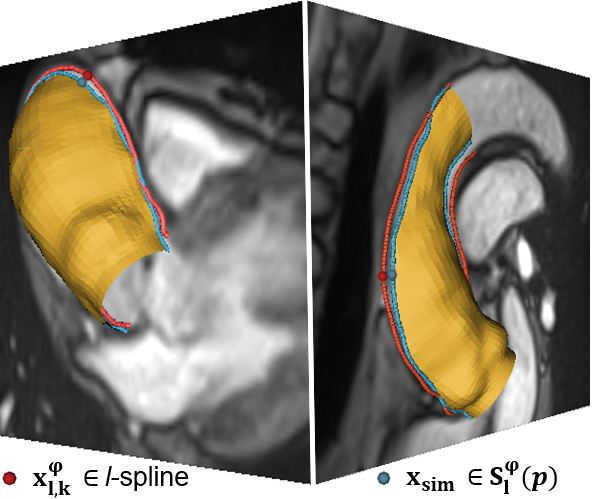}
\caption{Splines for two given cine-MRI sequences in a specific frame.}

\label{distance_cine}
\end{figure}

The whole calibration was performed only on the ascending aorta wall $\Gamma_{AAw}$, defined from the annulus to the plane perpendicular to the centerline at the level of the BCA ostium, 
where the experimental tests were performed. Each iteration of the calibration involved deriving a new zero-pressure model and then performing a structural analysis in which, in a time equivalent to one cardiac cycle, the load to reach the diastolic configuration was applied and thereafter 3 complete cardiac cycles were simulated.
The update of the parameters at each iteration was performed through a least squares optimisation using Levenberg-Marquardt (LM) method \cite{more1978levenberg} and minimizing the following loss function consisting of the $L_2$ norm:

\begin{equation}
f(\mathbf{p})= \sqrt{\sum_{\varphi}^{}\sum_{l=1}^{m}\sum_{k=1}^{n_{l}}\left| d^{\varphi}_{l,k}(\mathbf{p})\right|^{2}}
\label{cost function}
\end{equation}
where $m$=11 is the number of splines derived from the segmented aortic boundaries during each cardiac phase selected, $\mathbf{p}=[W_{X},W_{Y},W_{Z},K_{ST}]$ is the vector containing the 4 input parameters to be optimised, $n_{l}$ is the number of points defining the $l$-spline  and $d^{\varphi}_{l,k}$ is the nearest neighbour distance of each point $k$ belonging to the spline $l$ and all the points of the corresponding spline derived from the simulated model. In this way, for each point of the reference spline, we were searching for the nearest point in the second point cloud and computing the euclidean distance. In other words, if $\mathbf{x_{l,k}^{\varphi}}$ is the $k$-point of the $l$-spline and  $\mathbf{x_{sim}}$ is a point of the set $\mathbf{S_l^{\varphi}(p)}$ obtained from the intersection of the same planes of the cine-MRI and the 3D model,   as graphically shown in Fig. \ref{distance_cine}, the function $d^{\varphi}_{l,k}(\mathbf{p})$, used as error metric, was defined as:
\begin{equation}
d^{\varphi}_{l,k}(\mathbf{p})=d(\mathbf{x_{l,k}^{\varphi}},\mathbf{S_l^{\varphi}(p)})=\min_{\mathbf{x_{sim}}\in \mathbf{S_l^{\varphi}}(\mathbf{p)}}\left \| \mathbf{x_{l,k}^{\varphi}}- \mathbf{x_{sim}} \right \|
\label{distance}
\end{equation}

Regarding the resolution of LM optimisation, a 50 iterations threshold and the following ending criteria were imposed:
\begin{itemize}
\item parameters stability: $\mathrm{max}\left | h_i^{IT} / p_i^{IT} \right |<\epsilon_1$
\item loss function stability: $f(\mathbf{p}^{IT}) - f(\mathbf{p}^{IT-1})< \epsilon_2$
\item gradient convergence: $\mathrm{max}\left | \mathbf{J}^{IT^{T}} \mathbf{r}^{IT} \right |<\epsilon_3$
\end {itemize}

where the superscript $IT$ indicated the iteration of the calibration loop, $\mathbf{h}^{IT}$ the vector containing the updates of the 4 parameters, $\mathbf{J}^{IT}$ the Jacobian matrix, $\mathbf{r}^{IT}$ the vector with the residual distances. The threshold parameters used were $\epsilon_1$=10$^{-3}$, $\epsilon_2$=5$\cdot$10$^{-3}$m and $\epsilon_3$=10$^{-3}$. To quantify the effect of the parameters tuning and the annulus motion on the strain at the aortic wall, we performed and compared 3 additional simulations: the first by removing the heart motion on  $\Psi_{inlet}$ and with the non-calibrated Robin BCs; the second by introducing the heart motion effect but still on the model with the non-tuned parameters; the third by removing again the annulus motion but considering the parameter values of the Robin BCs at the end of the calibration procedure.
The structural simulations were executed using LS-Dyna. The calibration was run on a Dell Precision 7820 workstation with 2 16-cores Intel® Xeon Gold 5218 and 256 Gb RAM. 

\subsection{Parameter range and sensitivity analysis}
To build the baseline FE model, we assumed initially isotropic Robin BCs with values derived from literature \cite{gindre2016patient}: $\mathbf{p^1}$=[1,1,1,10$^5$ Pa/m].
The minimum value of the range for the three weights $W_X$, $W_Y$, $W_Z$ was set to 0 in case of no relationship between the spine and the aorta and external support due only to the surrounding soft tissue. In this particular configuration, the minimum for $K_{ST}$ was also assumed equal to 10$^4$ Pa/m, value below which the convergence was not achieved due to the insufficiently constrained model. As the stiffness of the Robin conditions increased, the model was more and more unable to expand during the cardiac cycle. Considering that, to obtain the maximum value of the parameters, we changed one parameter at a time starting from the initial guess. Assuming a difference of 1 mm in the ascending aorta mean diameter between diastole and systole \cite{stefanadis1990distensibility}, the maximum value of each range was identified when the new parameter value resulted in diameter variation smaller than this threshold. Thus, the maximum value of the ranges were 2.2, 1.3, 1.4 and 2$\times$10$^6$ Pa/m respectively for $W_X$, $W_Y$, $W_Z$ and $K_{ST}$. The $\boldsymbol{\varepsilon}$ vector containing the normalised updating values for building the Jacobian matrix $\mathbf{J}^{IT}$ was found through a sensitivity analysis of the convergence of the partial derivative to preserve the linear  approximation of the first order Taylor expansion \cite{levenberg1944method}. For each parameter of $\mathbf{p^1}$, we executed an iteration of the workflow using each time an increment value $\in \left\{ 0.1,0.05,0.02,0.01 \right\}$. The convergence of the partial derivatives was achieved with $\boldsymbol{\varepsilon}=[0.02, 0.01,0.01,0.02]$. 

To refine the robustness of the method and assess the sensitivity of the output based on the choice of initial guess, the Response Surface (RS) methodology \cite{khuri2010response} was used. A Design of Experiments based on 16 Design Points (DP$_{1}$,..., DP$_{16}$) obtained within the parameter ranges by Latin hypercube was prepared \cite{viana2010algorithm}. The cost function was computed for each of the 16 Design Points. The RS was calculated using the Genetic Aggregation method \cite{fiala2022practical} and cross-validated using leave-one-out, reporting the final root mean square error (RMSE$_\mathrm{{RS}}$) and the relative maximum absolute error (RMAE$_\mathrm{{RS}}$). The RS ability to predict the output was evaluated using the coefficient of determination (R$_{\mathrm{RS}}^{2}$).
The full calibration was then initialised again using 3 Design Points (DP$_{1}$, DP$_{2}$ and DP$_{3}$) for which the norm of the distance in the normalized ranges between them and $\mathbf{p^1}$ was the maximum.

\subsection{Strongly-coupled FSI}
Once the parameters were tuned, we performed a fully-coupled fluid-structure interaction analysis to examine the differences between the deformed structural domain derived from this simulation and the deformed aortic wall resulting from the calibration based on purely structural simulation. The same fluid-dynamic and structural models previously presented were used, as well as the same boundary conditions. In order to start the FSI simulation with the zero-pressure state, the solid domain was deformed by radial basis function (RBF) mesh morphing \cite{biancolini2012mesh}. Mesh morphing can exploit a motion imposed on a set of surface nodes acting as Source Points in order to perform a RBF interpolation in space that also changes the position of the volume nodes \cite{geronzi2021high}. In our case, we used all surface nodes of the original FE model derived from the segmentation in diastole as Source Points and we imposed a translation to each of them to achieve the computed zero-pressure state related to the calibrated model. Regarding the FSI coupling details, the maximum number of sub-iterations for the resolution was set at 50. Minimum and maximum time steps were equal to 1e-5 s and 5e-5 s, respectively. The FSI method used was the penalty-coupling algorithm. The comparison with the structural simulation was done by analysing the two deformed surface grids and reporting the maximum error assessed as maximum Euclidean distance (D$_{MAX_{\mathrm{FSI}}}$) between corresponding nodal positions  and maximum root mean square error (RMSE$_{\mathrm{FSI}}$) during the entire cardiac cycle.  Finally, calling $\overline{\mathbf{p}}$ the vector containing the calibrated parameters, a new cost function $f_{\mathrm{FSI}}(\overline{\mathbf{p}})$ was computed according to (\ref{cost function}) but using the deformed wall of the fully-coupled FSI simulation.

\begin{figure}[!b]
\centering
\includegraphics[width=75mm]{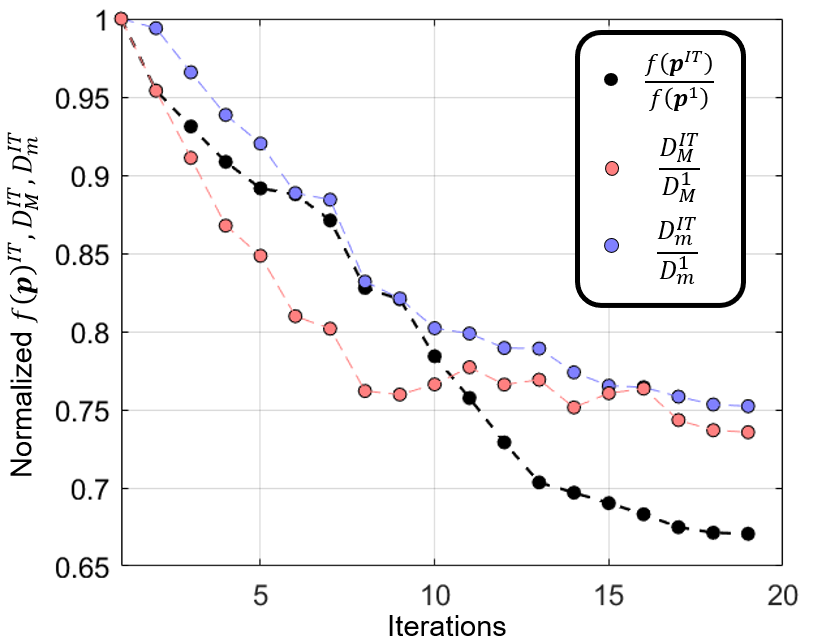}
\caption{Normalized loss function $f(\mathbf{p}^{IT})$, maximum $D_M^{IT}$ and mean $D_m^{IT}$ neighbour distance for each iteration of the LM optimisation.}
\label{fig5}
\end{figure}

\section{Results}
After a thorough construction of the fluid-dynamic and structural models, the calibration procedure using Levenberg-Marquardt was fully automated and consequently  guaranteed reproducibility without introducing any manual user error. 

For the last five iterations of the ICP alignment, negligible rigid displacements of less than 0.08 mm were observed.

The patient-specific diastolic pressure field $P_w(t_{diast}$) to compute the zero-pressure state was in a range between 70 mmHg and 73 mmHg depending on the aortic region.

\begin{figure*}
\centering
\includegraphics[width=162mm]{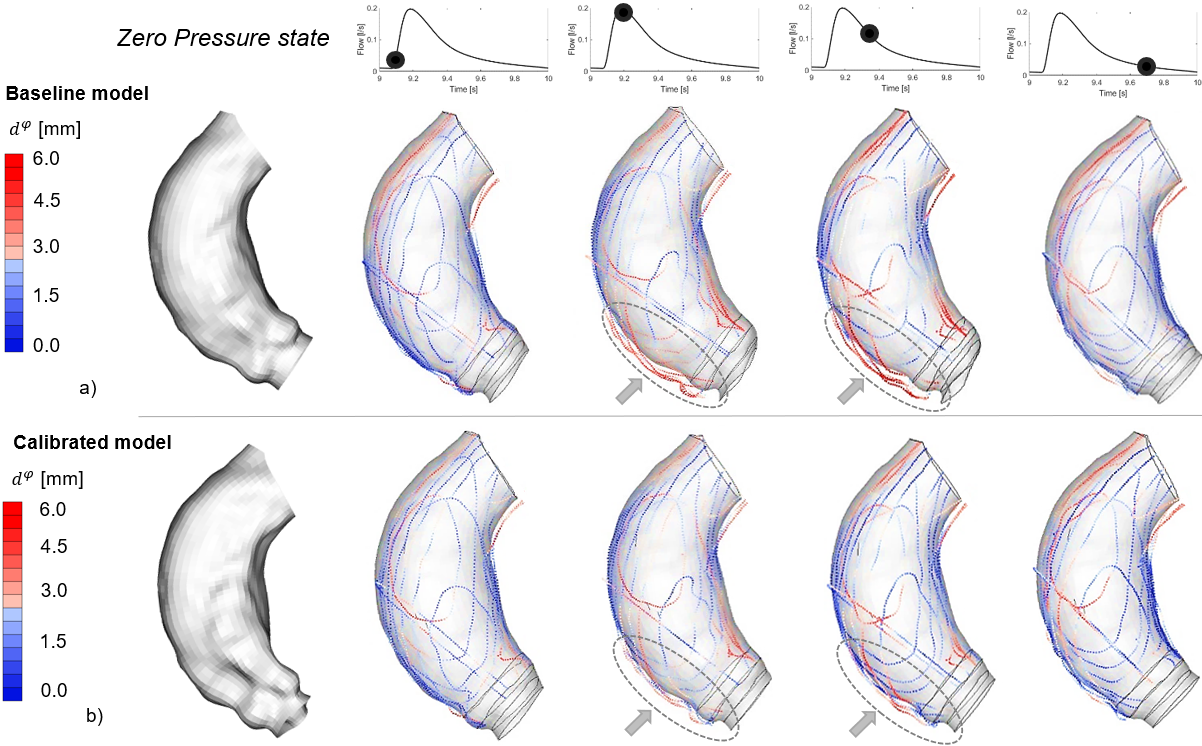}
\caption{Moving to the right: ascending part after the zero-pressure calculation, deformed vascular structure and error in term of neighbour distance on the cine-MRI-derived splines during the early systole, peak systole, late systole and early diastole for a) the baseline model above, b) the calibrated model below. The error is reduced mainly in the anterior-lateral zone of the initial ascending part during the peak and late-systolic frames.}
\label{fig8}
\end{figure*}

\begin{figure}[b]
\centering
\includegraphics[width=88mm]{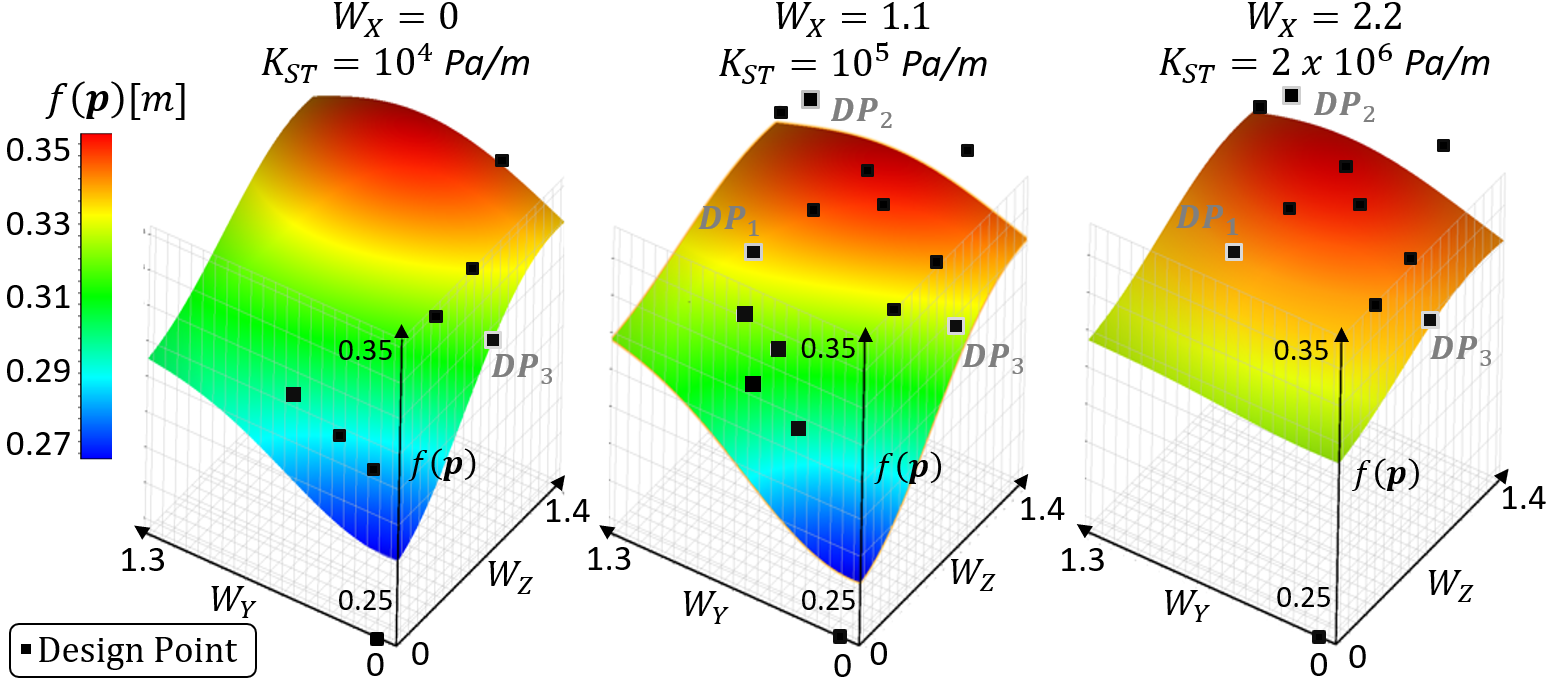}
\caption{Response Surface shown by fixing 3 values of $W_X$ and $K_{ST}$. DP$_1$, DP$_2$ and DP$_3$ were used to test the robustness of the method.}
\label{fig_rs}
\end{figure}

\begin{figure}
\centering
\includegraphics[width=84mm]{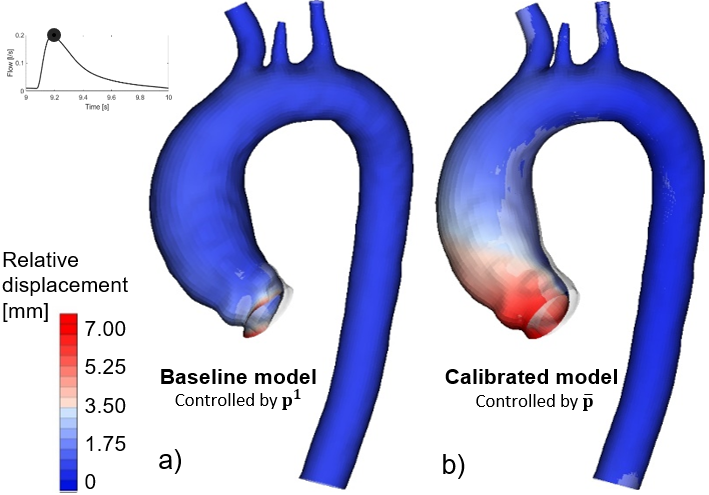}
\caption{Relative systolic-diastolic displacement contours for a) the case with the initial parameters guess and b) the calibrated model.}
\label{fig7}
\end{figure}

The calibration was successfully performed on the analysed patient and the loss function $f(\mathbf{p})$ was reduced by approximately 34\% after 19 iterations, going from 0.343 m to 0.227 m. The calculation took 32 hours in total. The algorithm stopped due to the second optimisation ending condition.  The value of the parameters governing the mechanical BCs at the end of the calibration was $\overline{\mathbf{p}}=[0.6, 0.02, 0.04, 1.5 \times 10^4$ Pa/m$]$. The evaluations of the cost function during the procedure, normalised with respect to the value of the first iteration, are shown in Fig. \ref{fig5}. In addition, for each iteration, the maximum and mean normalized neighbour distance between the splines,  $D_M^{IT}$ and $D_m^{IT}$ respectively, are reported in the same figure. These quantities are defined as the maximum and mean value of the vectors containing the evaluation of (\ref{distance}) on each point. The first decreased from 8.64 mm to 6.37 mm and the second from 2.24 mm to 1.83 mm. At the end of the calibration, the maximum error in terms of distance between the points belonging to the two sets of splines was  in the area of the sinotubular junction at the phase immediately following the systolic peak (Fig. \ref{fig8}) when the displacement of the annulus was more pronounced. In performing each LM iteration, the number of simulations to obtain the zero-pressure state varied from 4 to 15, in accordance with what reported in \cite{rausch2017augmented}. In general, as the spring stiffness decreased, a higher deformation corresponded and consequently a higher number of iterations to reach the convergence criterion was required. In fact, the baseline model controlled by the initial guess $\mathbf{p^1}$ required an average displacement on the ascending aorta of 1.1 mm to reach the unloaded configuration, whereas the final calibrated model, whose BCs were governed by $\overline{\mathbf{p}}$, having globally lower-value parameters, presented an average displacement of 5.4 mm. The Response Surface with the three Design Points used to verify the calibration results is reported in Fig. \ref{fig_rs}. Being a 4-parameters space, its 3D representation is obtained by fixing two of them. A good quality and prediction ability was ensured by R$_\mathrm{{RS}}^{2}$ = 0.998, RMSE$_\mathrm{{RS}}$ = 1.8 $\times 10^{-5} $ m and RMAE$_\mathrm{{RS}}$ = 2.91\%. $DP_{1}$ and $\mathrm{DP}_{3}$ returned the same minimum  $f(\overline{\mathbf{p}}_{\mathrm{DP}_{1}}) = f(\overline{\mathbf{p}}_{\mathrm{DP}_{3}}) = f(\overline{\mathbf{p}})=$ 0.227 m  after 21 and 16 iterations with $\overline{\mathbf{p}}_{\mathrm{DP}_{1}}=\overline{\mathbf{p}}_{\mathrm{DP}_{3}}=\overline{\mathbf{p}}$ while $\mathrm{DP}_{2}$, after 25 iterations, returned a new vector $\overline{\mathbf{p}}_{\mathrm{DP}_{2}}=[0.64, 0.04, 0.04, 1.4 \times 10^4 $ Pa/m$]$ for which $f(\overline{\mathbf{p}}_{\mathrm{DP}_{2}})=0.232$ m, 2.2\% higher than the minimum previously obtained.
The effect of the calibration on the aortic motion is shown in Fig. \ref{fig7}: the aortic displacement in systole compared to the late diastolic position in background is shown for the baseline model simulated with the initial guess  $\mathbf{p^1}$ and for the model at the end of the calibration controlled by $\overline{\mathbf{p}}$. In Fig. \ref{fig_strain}a, the strain contours are shown to highlight the calibration and the annulus motion effects. The maximum strain values are instead given in Table \ref{tab2}.

\begin{table}
\caption{Maximum von-mises equivalent strain: effect of calibration (C) and heart motion (HM).}
\setlength{\tabcolsep}{2pt}
\begin{tabular}{|p{27pt}|p{50pt}|p{50pt}|p{50pt}|p{50pt}|}
\hline
Phase & 
HM: no C: no &
HM: yes C: no &
HM: no C: yes &
HM: yes C: yes\\
\hline
\vspace{0.0005 in} Diastole & \vspace{0.0005 in}  0.051 &  \vspace{0.0005 in}  0.051 & \vspace{0.0005 in}  0.119 & \vspace{0.0005 in}  0.119 \\
\vspace{0.0005 in} Systole & \vspace{0.0005 in}    0.131 & \vspace{0.0005 in}   0.294 & \vspace{0.0005 in}    0.251 & \vspace{0.0005 in}    0.262 \\
\hline
\end{tabular}
\label{tab2}
\end{table}

Solving the fully-coupled fluid-structure interaction analysis required 108 hours. Comparing the displacement derived from strongly-coupled FSI and structural simulation (Fig. \ref{fig_strain}b), the maximum error between corresponding nodes $D_{MAX_{\mathrm{FSI}}}$ = 0.64 mm was detected for a node close to the annulus at the systolic peak. Precisely at this phase, the solutions of the two methods showed  differences in deformation near the lower part of the sinuses of Valsalva. RMSE$_{\mathrm{FSI}}$ was equal to 0.12 mm in late diastole and 0.19 mm at the systolic peak, the maximum value observed during the entire cardiac cycle. The cost function evaluated on the FSI deformed wall returned $f_{\mathrm{FSI}}(\overline{\mathbf{p}})$ = 0.237 m, value 4.4\% higher  than $f(\overline{\mathbf{p}})$.

\section{Discussion}
\subsection{Calibration and FSI results}

In this research, we presented a method to perform a patient-specific calibration of the parameters governing the mechanical boundary conditions of a high-fidelity thoracic aorta model. 

The included Robin BCs have two main advantages: from a clinical perspective, they restrict the aortic movement improving the correspondence with the motion derived from the images, whereas on the computational side, they affect the movement in areas typically left unconstrained (the wall) and relaxes portions often strongly constrained (the inlets and outlets) where the use of constraint equations for the cardiac motion  allows the aorta to expand and deform due to the pressure load.
The choice to tune the BCs by exploiting only  $\Gamma_{AAw}$  is due to two reasons: first, material properties data were available only for this region and second, the available cine-MRI images do not allow the accurate identification of displacements in the arch and descending aorta due to partial volume effect or to the absence of intersections with the considered aortic part. Moreover, the descending aorta motion is usually less pronounced and its displacement is often smaller than the resolution of the cine-MRI \cite{carrascosa2013thoracic}.

During each iterative step, an inverse problem was solved to obtain the zero-pressure state.
The diastolic pressure field applied in this simulation was consistent with the patient's blood pressure detected during the MRI acquisition. The maximum difference between each value of the pressure field and the measured pressure was 2 mmHg, aligned with the differences identified in literature \cite{chemla2021new} between aortic diastolic pressure and peripheral diastolic pressure. The unloaded configuration obtained was mainly deformed in the ascending aorta as here the stiffness values modelled in (\ref{stiffness}) were the weakest. The sets of springs connected to the three upper branches  to represent the effect of the upstream vasculature had the additional benefit of stabilizing the solution of the zero-pressure inverse problem, constraining even more the displacements of the nearby areas.

The threshold values for the ending criteria  related to the LM optimisation were not set very strictly for two reasons: first, if lower, these would significantly increase the number of iterations and excessively prolong the computation time. Second, at each evaluation of the cost function, the  residual error introduced by the zero-pressure inverse problem could potentially outweigh the improvement achieved by the LM evaluation itself. A strict threshold of 0.1 mm was set for the zero-pressure convergence, reducing the impact of computation errors on the final assessment of the cost function.

In each iteration of the workflow, the distance between the points belonging to the segmentation-derived splines and those to the simulation-derived splines, given by (\ref{distance}), varied during the cardiac cycle and the maximum error occurred in systole which is characterized by the highest vessel wall dilation and displacement \cite{greenfield1962relation}, as reflected by the frames sampling  $\varphi$ which is denser for the first frames in systole than for the last  in diastole \cite{plonek2018evaluation}.  The loss function considered all the frames $\varphi$ but the ones in systole had the strongest influence on reducing its value after changing the parameters, as can be observed from Fig. \ref{fig8}. It is difficult to reduce the contribution of the diastolic phase on the error because it is mainly due to segmentation bias which introduces a residual distance between the 3D computational model and the splines derived from cine-MRI. This error is mainly observed in the medial part at the end of the ascending aorta close to the pulmonary artery and in the antero-lateral zone where the vena cava is in contact with the aorta.

\begin{figure*}
\centering
\includegraphics[width=170mm]{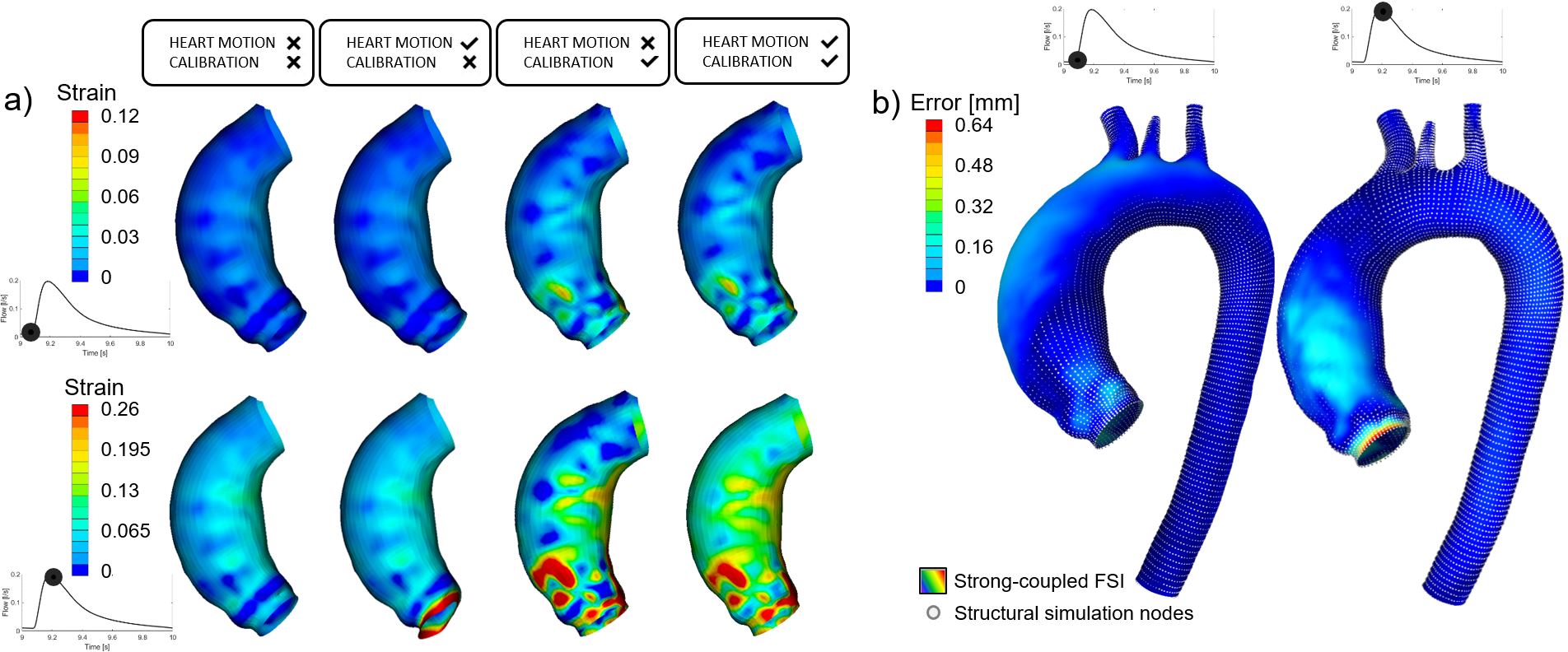}
\caption{a) Effect of imposed annulus motion and calibration on strain contours for the ascending aorta in late diastole (top) and at the systolic peak (down). b) Distance between each node of the wall in fully-coupled FSI simulation and the corresponding node in structural simulation.}
\label{fig_strain}
\end{figure*}

It is worth pointing out that $f(\mathbf{p})$  could not be reduced to 0 due to the limited control provided by only 4 variables; more parameters may probably be required to reduce the error. The inspection of the Response Surface (Fig. \ref{fig_rs}) suggests a minimum for low values of $W_Y$ and $W_Z$, consistent with the minimum achieved by the LM optimisation. Moreover, the repetition of the workflow using the 3 additional DPs as initial guesses helped to partially demonstrate its robustness in solving this highly non-linear problem.
Due to the choice of the initial guess, the average nodal displacement to reach the unloaded state for the proposed model with tuned Robin BCs was higher than for the baseline model, phenomenon already observed in the FSI models with and without the Robin BCs proposed by Baumler et al. \cite{baumler2020fluid}.
The comparison between the baseline and calibrated model in Fig. \ref{fig7} graphically shows a clear difference in terms of relative systolic-diastolic displacement after the calibration of the Robin BCs: the model based on $\mathbf{p^1}$ was strongly constrained and the motion applied at the level of the annulus was propagated exclusively to the nearest elements up to the central part of the Valsalva sinuses. In the calibrated case, on the other hand, the entire ascending aorta absorbed the impulse caused by the heart muscle.

Having made the kinematics of the vessel more consistent with the information derived from the cine-MRI sequences allows the model with the tuned BCs to reproduce the real displacement and deformation more faithfully than without calibrated parameters.  An improved fidelity in reproducing this behaviour results in a better strain assessment, potentially making the model more accurate in predicting the risk of rupture related to pathologies such as ascending aortic aneurysms.
Fig. \ref{fig_strain}a shows how the boundary conditions affect the strain of the model. In diastole, no difference is observed in the strain field when applying the heart motion. After the calibration, the parameters of the Robin BCs change and the strain increases as the zero-pressure state is more compressed.  Regarding the systole, compared to the baseline case without annulus movement, the heart motion induces a localised increase in strain exclusively for the elements near the annulus. Removing the heart motion but calibrating the 4 parameters of the Robin BCs makes the strain higher due to the new zero-pressure configuration. Strains are even higher if heart motion and calibration are combined \cite{beller2004role}.
Comparing the iso-topological surface meshes in structural and fully-coupled FSI simulation (Fig. \ref{fig_strain}b), differences are also evident in diastole. This occurs because the zero-pressure grid, whose nodal positions are the same for both methods, was derived using the diastolic pressure field computed with rigid-wall CFD simulation. In fact, deformable walls introduce slight variations in the pressure load, thereby modifying the final position achieved during diastole. However, the maximum difference between the two deformed walls occurs at the systolic peak. Although this may seem high  (0.64 mm), it can be observed from Fig. \ref{fig_strain}b that the major errors are all concentrated in the area close to the annulus. This is due to the control method of the annulus movement based on the constraint equations proposed in (\ref{dummy}) through which we do not impose a displacement to every single node of $\Psi_{inlet}$ but control the overall behaviour of the entire set of nodes with respect to the dummy node. If the two simulations were compared exclusively in the calibration domain $\Gamma_{AAw}$, where the first elements close to the annulus were excluded, as shown in Fig. \ref{fig3}b, the maximum error in systole is reduced to 0.24 mm. This difference between the two methods is in agreement with what was reported in \cite{lin2017fluid}. The value of the cost function $f_{\mathrm{FSI}}(\overline{\mathbf{p}})$ demonstrates that calibrating the model using a pressure field and structural simulations yields tuned parameters that ensure an improved fidelity even for the FSI model. Moireau et al. \cite{moireau2013sequential} correctly highlighted the computational cost associated with calibration methods based on fluid-structure simulations aiming at minimizing the discrepancy between the computational model and image-derived information. Our approach stands as a viable alternative, since it exploits the decoupling of the physics, providing a huge gain in computational time.

Given that current imaging techniques do not usually allow a simple in-vivo estimation of the tissue properties of the aorta \cite{cosentino2019role}, we preferred to show the consistency of the calibration workflow on a model with aneurysm derived from a patient who underwent surgical repair. Nevertheless, the same procedure could be used to treat healthy aortas by adding new material-related parameters.  The proper execution of the workflow is then independent of the valvular inlet type: a different shape would mainly vary the CFD-derived wall pressure field $P_w(t)$, especially in the ascending aorta. The calibration here proposed could be similarly extended to the abdominal aorta, to the heart or the diaphragm for motion compensation in cancer radiation therapies \cite{ladjal2013appropriate}.
Finally, this procedure could be repeated with every image set that considers dynamic acquisitions, such as multiphase-CT or even 4D-Flow MRI. These techniques indeed provide information and time-resolved data on the aortic motion even allowing the description of phenomena such as the rotation of the aorta at the level of the annulus.  In any case, we consider cine-MRI data as a good compromise between  spatial resolution and image quality of CT scan and good temporal accuracy of 4D-Flow MRI \cite{pelc1991phase}. Shidhore et al. \cite{shidhore2022estimating} used 4D Ultrasounds images to perform some stiffness estimations of the Robin BCs for three macro-areas of a murine aorta FSI model. However, 4D ultrasound imaging is not currently part of the clinical practice for the analysis of ascending aortic aneurysms. While 2D ultrasound acquisitions are occasionally employed for screening purposes, they rely on user-dependent probe positions, making accurate calibrations challenging.

\subsection{Limitations and future work}
This work presents some limitations that need to be mentioned.
Firstly, the procedure is applied exclusively to one patient. It would be interesting to extend it to other cases to evaluate if the  obtained stiffness values are consistent and investigate a potential inter-patient variability. Secondly, the mechanical boundary conditions were controlled by only 4 parameters related to the stiffness of the Robin BCs. Further investigations are required to assess if additional parameters could increase the accuracy and fidelity of the model. We have then assumed the damping factor $\boldsymbol{\eta}$ constant and  equal in the three directions of space. In addition, a linear relationship that adjusts the stiffness of the springs according to the distance to the spine was assumed; alternative mathematical formulations should be tested to assess their impact on the outcomes. Concerning the TA model, we did not consider the effect of the coronary arteries and the 3D valvular shape. We impose then constant thickness and isotropic material for the entire vessel. Nevertheless, this does not alter the way the procedure works.  Furthermore, we did not take into account the self-balancing residual stress in the unloaded state \cite{azeloglu2008heterogeneous} and the different types of material properties related to the  intima, media and adventitia layers \cite{holzapfel2010modelling}. When extending this method to multiple subjects, attention should be paid to the parameters chosen for the LM algorithm to provide viable solutions. 

Improving the model fidelity by including morphological and functional aspects could enhance the understanding of the complex biological process related to the aneurysm growth and rupture \cite{souche2022high}. Extending this workflow to multiple patients, however, requires automation of some still laborious and time-consuming steps such as image segmentations and model generation. By addressing the aforementioned  limitations, this procedure could be used to return a patient-specific model able to provide accurate aneurysm risk assessments. This parameter-based model could also allow the realisation of a Digital Twin of the aorta \cite{coorey2021health} enabling real-time simulation of vascular replacement prostheses for the ascending aorta \cite{spadaccio2016old} and analyzing  thoracic endovascular procedures such as wire insertion for trans-catheter aortic valve replacement \cite {auricchio2014simulation}.

\section{Conclusion}

In this study, we presented a novel method for calibrating 4 parameters managing the mechanical boundary conditions of a high-fidelity thoracic aorta model, accounting for the motion imposed by the heart on the aorta at the level of the annulus. The combination of a hyper-elastic material based on experimental data with an iterative zero-pressure calculation could allow to more faithfully estimate bio-mechanical parameters. This is particularly important for understanding the progression of pathologies like aneurysms and aortic dissections, where the wall conditions play a crucial role.
Although further extensions are required, the application of the calibration procedure to a patient with an aneurysm demonstrated the possibility of improving the model fidelity, achieving a closer correspondence with information derived from the available image data. The procedure presented in this work has to be tested on a large dataset with the long-term goal of delivering accurate and reliable patient-specific analyses crucial for cardiovascular disease risk assessment and personalized therapeutic solutions.

\appendices

\section*{Acknowledgment}

The research received funding from the European Union’s Horizon 2020 programme under the Marie Skłodowska-Curie grant agreement No 859836, MeDiTATe.


\end{document}